\documentclass[twocolumn,superscriptaddress,showpacs,preprintnumbers,amsmath,amssymb,aps,prc]{revtex4}

\usepackage{amsmath}
\usepackage{graphicx}
\usepackage{amssymb}
\usepackage{bm}
\usepackage{color}

\begin{document}

\title{Gamow-Teller transitions from high-spin isomers in $N=Z$ nuclei}

\author{H. Z. Liang}
 \affiliation{RIKEN Nishina Center, Wako 351-0198, Japan}
 \affiliation{Department of Physics, Graduate School of Science, The University of Tokyo, Tokyo 113-0033, Japan}

\author{H. Sagawa}
  \affiliation{RIKEN Nishina Center, Wako 351-0198, Japan}
  \affiliation{Center for Mathematics and Physics, University of Aizu, Aizu-Wakamatsu, Fukushima 965-8560, Japan}

\author{M. Sasano}
  \affiliation{RIKEN Nishina Center, Wako 351-0198, Japan}

\author{T. Suzuki}
  \affiliation{Department of Physics, College of Humanities and Science, Nihon University, Tokyo 156-8550, Japan}
  \affiliation{National Astronomical Observatory of Japan, Mitaka, Tokyo 181-8588, Japan}

\author{M. Honma}
  \affiliation{Center for Mathematics and Physics, University of Aizu, Aizu-Wakamatsu, Fukushima 965-8560, Japan}

\begin{abstract}
Gamow-Teller (GT) transitions from high-spin isomers are studied using the sum-rule approach and the shell model.
The GT transition strengths from the high-spin isomeric states show a stronger collectivity than those from the ground states in two $N=Z$ nuclei, $^{52}$Fe and $^{94}$Ag.
It is argued that the spin-up and spin-down Fermi spheres involved in the GT transitions from the high-spin isomeric states play important roles.
These Fermi spheres are analogous to the isospin-up and isospin-down Fermi spheres for the GT transitions from the ground states in $N>Z$ nuclei and create a strong collectivity.
\end{abstract}


\maketitle
\section{Introduction}

Studies on Gamow-Teller (GT) transitions have provided essential knowledge about nuclei.
Transition operators ${\vec \sigma} t_{\pm}$ cause only rotation in the spin-isospin space and do not affect spatial wave functions \cite{Bohr1969}.
Due to the time-odd and isovector nature of GT transitions, a study of such a transition mode reveals specific information that cannot be obtained from other studies.
Because GT transition operators are also involved in the matrix elements that describe the weak-interaction processes in nuclei, studies on GT transitions are crucial to establish nuclear models to predict the rates of weak reactions in astrophysical phenomena \cite{Langanke2003, Niu2013} and in exotic weak processes such as double-$\beta$ decays \cite{Engel2017}.

For stable medium-heavy or heavy nuclei with neutron excess, it is well known that the GT transition mode exhibits a collective resonance called the GT giant resonance (GTGR) due to strong repulsive residual interactions in the particle-hole ($ph$) channel.
There exists a very useful sum rule for evaluating the behavior of the GTGR \cite{Ikeda1963}:
Regarding nuclei as non-relativistic systems consisting of nucleons, the difference between the non-energy-weighted summed strengths of the GT transitions in two opposite directions in the isospin space, the $\beta^-$ and $\beta^+$ directions, is written as
\begin{equation}
S_--S_+\equiv\sum_{f,\nu}|\langle f|\hat{O}^{-}_{\nu}|i\rangle|^2-
\sum_{f,\nu}|\langle f|\hat{O}^{+}_{\nu}|i\rangle|^2=3(N-Z)\,,
\end{equation}
where $N$ and $Z$ are the neutron and proton numbers of the initial nucleus, respectively.
See the definitions of $\hat{O}^{\pm}_{\nu}$ in Eq.~(\ref{ope-GT}) below.

Collectivity can be evaluated by how much strength is included in the GTGR compared to this sum-rule value.
Empirically, such a comparison has shown that about $60\%$ of the total strength lies in the energy region up to the GTGR, while the remaining $40\%$ is attributed to either the mixing effect of high-order configurations beyond the mean-field description of nuclei or the coupling between the nucleon and the $\Delta$-particle degrees of freedom \cite{Gaarde1981, Wakasa1997, Yako2005, Ichimura2006}.

One limitation of the GT sum rule is that its value is always zero for $N=Z$ nuclei.
However, this does not mean that these nuclei do not exhibit GTGR.
Consequently, evaluating the collectivity of the GT transitions in $N=Z$ nuclei is much more difficult.
The fundamental difference between $N=Z$ and $N>Z$ nuclei is that the isospin degree of freedom is averaged out in $N=Z$ nuclei, but is asymmetric in $N>Z$ nuclei.
In the case of $N>Z$ nuclei, GT transitions occur predominantly through $ph$ excitations, where holes are created in the larger neutron Fermi sphere and particles are created outside the smaller proton Fermi sphere \cite{Liang2008, Bai2009}.
In the case of $N=Z$ nuclei, this distinction between proton particles and neutron holes in terms of the isospin space is inappropriate, because the Fermi spheres of protons and neutrons are equivalent in the isospin space.
Therefore, in the GT transitions of these nuclei, the $ph$ and particle-particle ($pp$) excitations usually coexist due to the melting of the Fermi spheres \cite{Bai2013, Niu2013b}.
However, as discussed below, if the initial state of an $N=Z$ nucleus has a high spin, the GT transitions can be well described as the $ph$ excitations in the spin space.

In this paper, we will deduce a new GT sum rule for high-spin isomeric (HSI) states in $N=Z$ nuclei and discuss the collectivity of the GT transitions in these nuclei.
This theoretical paper is stimulated by an existing experimental project, which aims to measure the strengths of the GT transitions from the $I^\pi = 12^{+}$ isomeric state of the unstable nucleus $^{52}$Fe \cite{Yakoexp}.
In the stable region, $N=Z$ nuclei are limited within the mass region up to the $sd$-shell (i.e., $^{40}$Ca at the most).
With the advent of the worldwide radioactive-ion-beam facilities, a hot research topic is to provide heavier $N=Z$ nuclei.
Currently, experimentalists are planning to obtain the first experimental data of the GT transition strengths on such high-spin states in unstable $N=Z$ nuclei.

Herein we introduce a new concept: the spin Fermi sphere.
The spin-up and spin-down particles make separate Fermi spheres.
The HSI states in $N=Z$ nuclei can be regarded as nuclei with two different Fermi spheres for the spin-up and spin-down components, where protons and neutrons are filled symmetrically with respect to the isospin degree of freedom.
Assuming the spin-up component has more constituents than the spin-down one, a GT transition creates a hole in the Fermi sphere of the spin-up component and a particle outside the Fermi sphere of the spin-down component.
In contrast to $N>Z$ nuclei in the $0^+$ ground states, the roles of the spin and isospin degrees of freedom are interchanged in the definition of the Fermi spheres of the two components comprising the nuclei.
If the $ph$ residual interaction itself is the same as that in normal $N>Z$ nuclei, the GT transition from the HSI states in $N=Z$ nuclei should exhibit a highly excited and strong GTGR through the $ph$ residual interaction defined by these Fermi spheres characterized by the spin degree of freedom.

The above argument is simplified to explain the basic ideas of the new sum rule and the GTGR from the HSI states in $N=Z$ nuclei.
In reality, the quantization axis of spin must be defined to the intrinsic frame of nuclei because all the spin directions are averaged out in the laboratory frame.
It is possible to take the quantization axis along the direction that the nuclei are elongated when the nuclei are substantially deformed.

The aim of this paper is to clarify a new degree of collectivity in terms of the spin Fermi spheres using the GT sum rule from the HSI states and the widths of the calculated transition strength distributions by the large-space shell models.
We apply the new GT sum rule to the $I^\pi = 12^+$ and $21^+$ isomeric states of $^{52}$Fe and $^{94}$Ag, respectively, which are considered to be substantially deformed.
In addition, coupling of the spins to the orbital angular momenta must be considered in a general manner within the models based on deformation.
In this paper, we also perform the shell-model calculations for the GT transitions in these two $N=Z$ nuclei with HSI states and compare the results with the predictions based on the sum rule.

This paper is organized as follows:
Section~\ref{sec:Framework} is devoted to the study of the sum rule for GT transitions from HSI states.
Shell model calculations for the $12^+$ and $21^+$ isomeric states of $^{52}$Fe and $^{94}$Ag are discussed in terms of the sum rule values in Sections~\ref{sec:Fe52} and \ref{sec:Ag94}, respectively.
A summary is given in Section~\ref{sec:Summary}.

\section{Gamow-Teller transitions in deformed nuclei}\label{sec:Framework}

We consider the Gamow-Teller transition operators in the intrinsic frame of deformed nuclei,
\begin{equation}\label{ope-GT}
 \hat{O}^{\pm}_{\nu}=\sum_{\alpha}\sigma_{\nu}(\alpha)t_{\pm}(\alpha)\,,
\end{equation}
where $\nu=\pm 1,\, 0$ and $\alpha$ is the index of nucleons.
We evaluate the sum rules for the operators $ \hat{O}_{\pm1}$,
\begin{equation}\label{ope}
  S(\sigma_{\pm1}t_{\pm})=\sum_f|\langle f|\sum_{\alpha}\sigma_{\pm1}(\alpha)t_{\pm}(\alpha)|i\rangle|^ 2\,.
\end{equation}
Simple analytic formulas can be driven for the combinations of sum rules in Eq.~(\ref{ope}),
\begin{eqnarray}
    \Delta S(t_{-})&\equiv& S(\sigma_{-1}t_{-})-S(\sigma_{+1}t_{-})   \nonumber \\
    &=& 2\sum_{\alpha}\langle i|\sigma_0(\alpha)t_+(\alpha)t_-(\alpha)|i\rangle\,,  \label{summ}\\
    \Delta S(t_{+})&\equiv& S(\sigma_{-1}t_{+})-S(\sigma_{+1}t_{+})  \nonumber \\
    &=& 2\sum_{\alpha}\langle i|\sigma_0(\alpha)t_-(\alpha)t_+(\alpha)|i\rangle\,.
    \label{sump}
\end{eqnarray}
It should be noted that Eq.~(\ref{summ}) corresponds to the GT transitions from a nucleus $(N,\,Z)$ to its isobaric neighboring nucleus $(N-1,\,Z+1)$, while Eq.~(\ref{sump}) corresponds to those from $(N,\,Z)$ to $(N+1,\,Z-1)$.
Thus, combinations of Eqs.~(\ref{summ}) and (\ref{sump}) become model-independent sum rules, which are expressed as
\begin{eqnarray}
\Delta S(t_{-})-\Delta S(t_{+})&=& 4\sum_{\alpha}\langle i|\sigma_0(\alpha)t_z(\alpha)|i\rangle   \nonumber \\
&=& 2( \langle S_n \rangle -  \langle S_p \rangle)\,,\\
\Delta S(t_{-})+\Delta S(t_{+})&=& 2\sum_{\alpha}\langle i|\sigma_0(\alpha)|i\rangle   \nonumber \\
&=& 2( \langle S_n \rangle + \langle S_p \rangle)\,,
\end{eqnarray}
where $\langle S_{n(p)}\rangle\equiv\sum_{\alpha \in n(p)}\langle i | \sigma_0(\alpha)| i \rangle$ is the spin polarization along the symmetry axis.

The sum-rule values can be obtained explicitly by the single-particle wave functions with Nilsson quantum numbers $[Nn_3\Lambda\Omega]$, where $N$ is the major oscillator quantum number.
Defined along the symmetry axis, $n_3$ is the number of quanta in the oscillation, $\Lambda$ is the projection of the orbital angular momentum, $\Sigma=\pm1/2$ is the spin projection, and $\Omega$ is the projection of the total angular momentum $\Omega=\Lambda+\Sigma$, respectively.
The Nilsson state $[Nn_3\Lambda\Omega]$ can be expanded by spherical $jj$-coupling wave functions $|j\Omega\rangle$ labeled by the total angular momentum $j$ and its projection on the symmetry axis $\Omega$.
These quantum numbers are relevant for the maximum aligned states along the symmetry axis such as $[Nn_3\Lambda\Omega]=[303\frac{7}{2}]$ and $[404\frac{9}{2}]$, which are equivalent to $|j\Omega\rangle=|\frac{7}{2}\frac{7}{2}\rangle$ and $|\frac{9}{2}\frac{9}{2}\rangle$, respectively.

First, we consider the spin operators $\sigma_{\pm1}$.
The sum of the single-particle transitions for $\sigma_{-1}$ gives
\begin{equation}\label{mem}
  \sum_{j'}|\langle j' (\Omega-1)|\sigma_{-1}|j\Omega\rangle|^2=\frac{j+\Omega}{j}\,,
\end{equation}
for the so-called intruder states with $j=l+1/2$.
In contrast, the matrix element for $\sigma_{+1}$ gives
\begin{equation}\label{mep}
  \sum_{j'}|\langle j' (\Omega+1)|\sigma_{+1}|j\Omega\rangle|^2=\frac{j-\Omega}{j}\,.
\end{equation}
The difference between Eqs.~(\ref{mem}) and  (\ref{mep}) is equal to twice the expectation value for operator $S$ of spin polarization, which reads
\begin{equation}\label{spin-me}
 \langle S\rangle = \sum_{\alpha} \frac{\Omega_{\alpha}}{j_{\alpha}}\,,
\end{equation}
for the single-$j$ configuration.
In addition, for operator $\sigma_0$, the matrix element is given as
\begin{equation}\label{me0}
  \sum_{j'}|\langle j' \Omega|\sigma_{0}|j\Omega\rangle|^2=1\,.
\end{equation}

The angular-momentum projection of intrinsic wave function $\Psi_{K}$ with $K=\sum_{i\in \alpha}\Omega_i$ to laboratory wave function $\Phi_{KIM}$ can be expressed  by a $D$-function as \cite{Bohr1975}
\begin{align}\label{lab-wf}
  \Phi_{KIM}  = &  \left(\frac{2I+1}{16\pi^2}\right)^{1/2} \nonumber \\
  &\times
  \left(\Psi_{K}(q)D^I_{MK}(\omega)+(-)^{I+K}\Psi_{\bar{K}}(q)D^I_{M-K}(\omega)\right)\,,
\end{align}
where $q$ is the intrinsic coordinates, which include spin variables.
$\omega=(\phi,\,\theta,\,\psi)$ are the orientation angles of the intrinsic wave function to the laboratory frame, and $\Psi_{\bar{K}}(q)$ is a time-reversed state $\Psi_{\bar{K}}(q)=T\Psi_{K}(q)$.
The GT transition operator $\hat{O}_{\rm GT}(1\mu)$ in the laboratory frame is obtained from the intrinsic operator~(\ref{ope-GT}) by a transformation
\begin{equation}\label{lab-GT}
  \hat{O}_{\rm GT}(1\mu)=\sum_{\nu}\hat{O}_{\nu}D^1_{\mu\nu}(\omega)\,.
\end{equation}
The reduced matrix element in the laboratory frame is then evaluated as
\begin{equation}\label{lab-me}
  \langle K'I'||\hat{O}_{\rm GT}||KI\rangle =
  (2I+1)^{1/2}\langle IK 1\Delta K|I'K'\rangle\langle K'|| \hat{O}||K\rangle\,,
\end{equation}
where $I=K$ and $K'=K+\Delta K$.
The values of the squares of the Clebsch-Gordan coefficients in Eq.~(\ref{lab-me}) are tabulated in Table~\ref{tab-CG}.

\begin{table}
  \caption{Clebsch-Gordan coefficients for the projection of the intrinsic frame to the laboratory frame of the GT transition matrix elements.}\label{tab-CG}
\begin{ruledtabular}
\begin{tabular}{cccc}
                 &  $I'=I-1$             & $I'=I$          &   $I'=I+1$               \\ \hline
  $\Delta K=-1$  &  ${(2I-1)}/{(2I+1)}$  & ${1}/{(I+1)}$ & ${1}/{[(2I+1)(I+1)]}$  \\
  $\Delta K=0$   & ---                   & ${I}/{(I+1)}$ & ${1}/{(I+1)}$          \\
  $\Delta K=+1$  & ---                   & ---             & 1                        \\
\end{tabular}
\end{ruledtabular}
\end{table}

\section{$I^{\pi}=12^+$ isomer in $^{52}$Fe}\label{sec:Fe52}

Nucleus $^{52}$Fe has an isomeric state at excitation energy $E=6.96$~MeV with spin-parity $I^\pi=12^{+}$ and half-life $T_{1/2}=45.9(6)$~s.
Experimentally, this $12^+$ state can decay to two $8^+$ states at $6.36$ and $6.49$~MeV, with the decay energies $0.60$ and $0.47$~MeV and the $B(E4)$ values $0.00046(22)$ and $0.0033(16)$~W.u., respectively \cite{ENSDF}.
This state can be considered as a four quasi-particle (qp) state of protons and neutrons in the prolate deformed potential.  The proton (neutron) configuration can be specified by the single-particle states with the Nilsson quantum numbers
$[303 \frac{7}{2}]_{\pi(\nu)}$ and $[312 \frac{5}{2}]_{\pi(\nu)}$.
For this 4-qp configuration, the GT matrix elements~(\ref{mem}), (\ref{me0}), and (\ref{mep}) in the intrinsic frame become
\begin{align}\label{int-me}
  {2(4j-1)}/{j} \qquad &\mbox{for} \qquad \Delta K=-1\,,  \nonumber \\
  4 \qquad &\mbox{for} \qquad  \Delta K=0\,, \nonumber \\
  {2}/{j} \qquad &\mbox{for} \qquad \Delta K=+1\,,
\end{align}
with $j=7/2$, respectively.
The values in Eq.~(\ref{int-me}) are obtained by assuming that the wave functions  $[Nn_3\Lambda\Omega]=[303 \frac{7}{2}]_{\pi(\nu)}$ and $[312 \frac{5}{2}]_{\pi(\nu)}$ have dominant contributions of $|j\Omega\rangle=|\frac{7}{2}\frac{7}{2}\rangle$ and $|\frac{7}{2}\frac{5}{2}\rangle$ orbits, respectively.
This assumption is quite reasonable for a large prolate deformation with $\beta_2 \sim0.3$.

\begin{table}
  \caption{$B_{\mathrm{GT}}$ strengths for the transitions $I\rightarrow I'$ with the 4-qp configuration of protons and neurons  $[Nn_3\Lambda\Omega]=[N0N(\Lambda+1/2)]_{\nu(\pi)}$ and $[N1(N-1)(\Lambda+1/2)]_{\nu(\pi)}$.
  Sum values in the last line are evaluated with $I=12$ and $j=7/2$.}\label{tab-GT}
\begin{ruledtabular}
\begin{tabular}{cccc}
      &  $I'=I-1$    &  $I'=I$  & $I'=I+1$  \\ \hline

  $\Delta K=-1$  &  $\frac{2(4j-1)(2I-1)}{j(2I+1)}$ &
                    $\frac{2(4j-1)}{j(I+1)}$ &
                    $\frac{2(4j-1)}{j(2I+1)(I+1)}$  \\
  $\Delta K=0   $   &---  &  $\frac{4I}{I+1}$ & $\frac{4}{I+1}$   \\
  $\Delta K=+1$¡¡   & ---   &   ---& $\frac{2}{j}$ \\\hline
  sum  & 6.83 & 4.26 & 0.90   \\
\end{tabular}
\end{ruledtabular}
\end{table}

The non-energy-weighted summed GT transition strengths for different $I'$ final states are tabulated in Table~\ref{tab-GT}.
The GT transition of $I\rightarrow I'=I-1$ is about $56.9\%$ of the total strength $B_{\mathrm{GT}}=12$.
The transition of $I\rightarrow I'=I$ holds about $35.5\%$ of the total strength, while that of $I\rightarrow I'=I+1$ is rather weak with $7.5\%$ of the total strength.

Microscopically, shell-model calculations of the GT transitions from the $I^\pi=12^+$ isomeric state in $^{52}$Fe are performed with a modern effective interaction GXPF1J \cite{Honma2005} in the $pf$-shell model space.
The $12^+$ state is found at $E=6.75$~MeV, which is slightly lower than the observed value.
The shell-model configuration of the $I^\pi=12^+$ state is dominated by $61\%$ of the $(f_{7/2}^2)_{\nu}(f_{7/2}^2)_{\pi}$ configuration with $6.4\%$ mixing of $(f_{7/2}f_{5/2})_{\nu}(f_{7/2}^2)_{\pi}$ or $(f_{7/2}^2)_{\nu}(f_{7/2}f_{5/2})_{\pi}$.
The other configurations involve the $p_{3/2}$ orbit and have the seniorities more than four.
For the decay scheme, the present shell-model calculations show this $12^+$ state decays to two $8^+$ states at $5.95$ and $6.33$~MeV.
The decay energies read $0.80$ and $0.42$~MeV, and the $B(E4)$ values with the bare charges read $0.0082$ and $0.036$~W.u., respectively.
It is seen that the two shell-model transition rates as well as the empirical ones are very much quenched to guarantee the isomeric nature of the $12^+$ state.
Quantitatively, the former $B(E4)$ value is comparable with its experimental counterpart, although the later one is larger than the experimental one.

\begin{figure}
    \includegraphics[width=8cm]{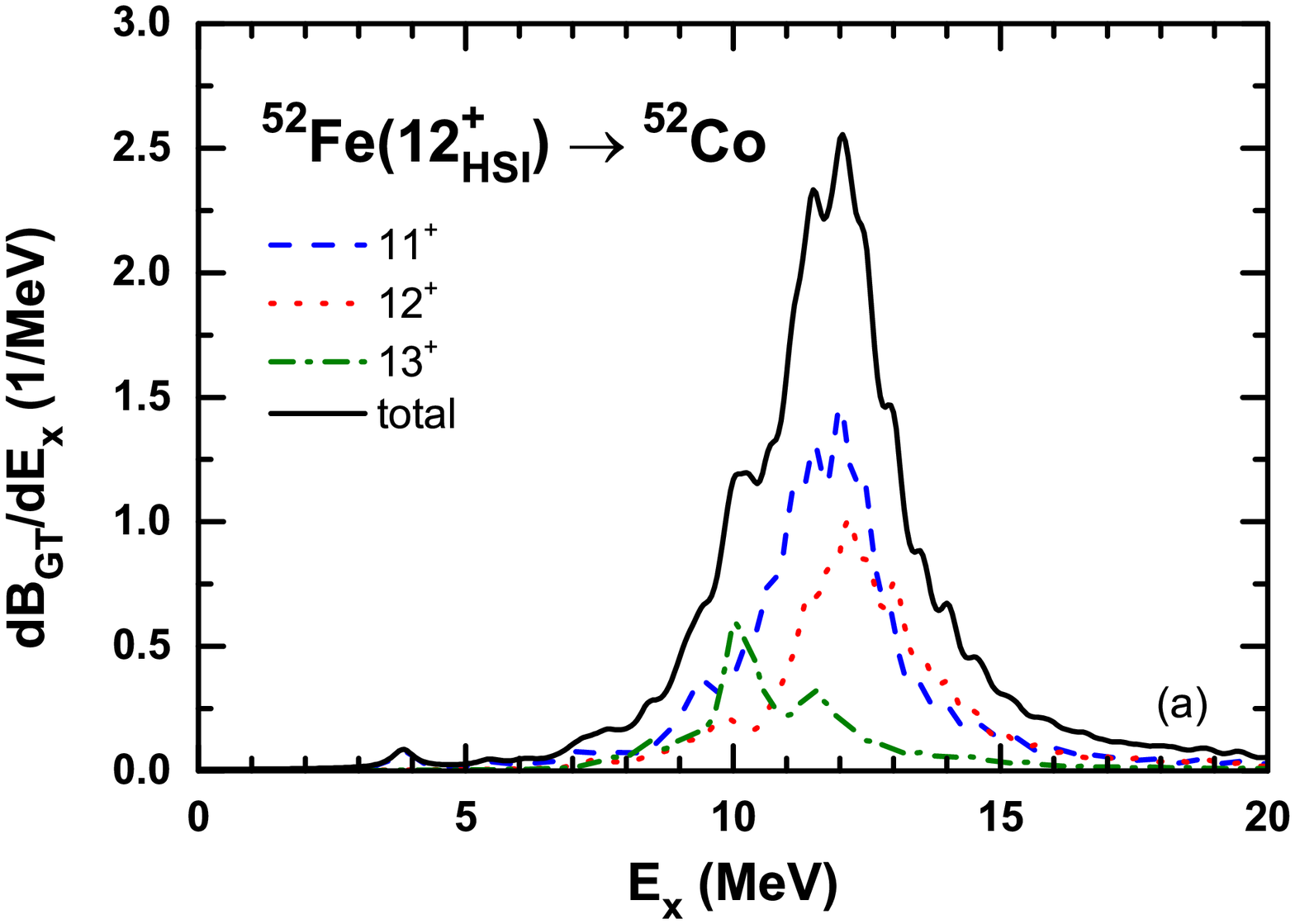}\\
    \includegraphics[width=8cm]{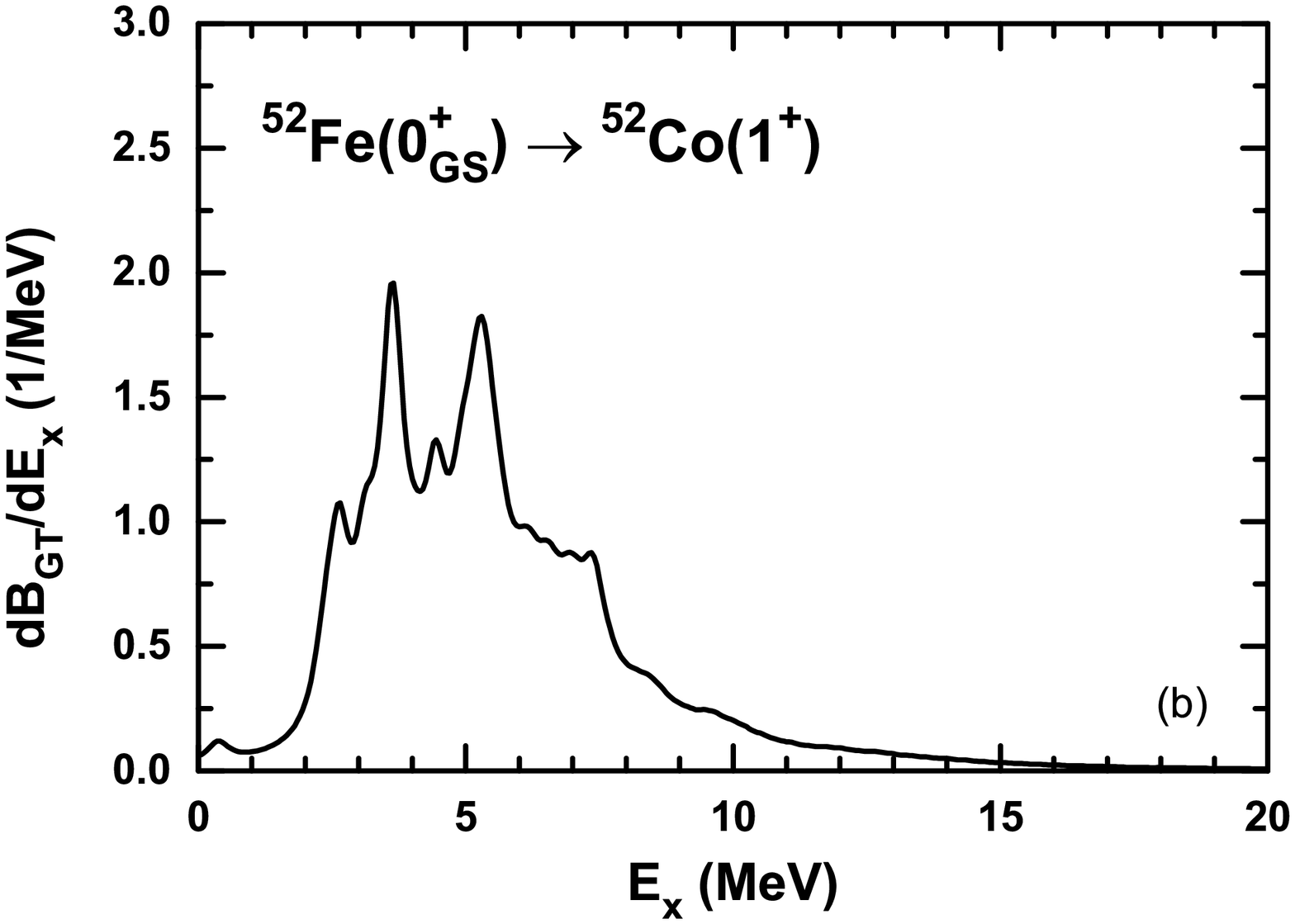}
    \caption{(Color online) (Upper) Gamow-Teller (GT) transition strength distributions from the high-spin isomeric state $I^{\pi}=12^+$ in $^{52}$Fe calculated by the shell model in the full $pf$-shell model space with the GXPF1J \cite{Honma2005} effective interaction.
    Transition strengths are averaged by a Lorentzian weighting function with a width of $0.5$~MeV.
    Excitation energy is defined from the ground state of the daughter nucleus $^{52}$Co.
    No quenching factor is adopted here.
    (Lower) Same as the upper panel but for the GT transition strength distributions from the ground state $I^{\pi}=0^+$ in $^{52}$Fe.
    \label{fig:fe52-gt}}
\end{figure}

The upper panel of Fig.~\ref{fig:fe52-gt} shows the corresponding GT transition strength distributions from the $I^{\pi}=12^+$ HSI state in $^{52}$Fe.
The calculated transition strengths are averaged by a Lorentzian weighting function with a width of $0.5$~MeV, and the excitation energy $E_x$ is defined from the ground state of the daughter nucleus $^{52}$Co.
Note that in the present results, no quenching factor is adopted, as we basically have no knowledge about the proper quenching factor for the GT transitions from the HSI states.
It is seen that the peaks of the $I'^\pi=11^+$ and $12^+$ components are located at $E_x=12.0$ and $12.2$~MeV, respectively, while that of the $I'^\pi=13^+$ component is at a lower energy $E_x=10.0$~MeV.
The summed transition strengths are $4.47$, $3.11$, and $1.44$ for the $I'=11$, $12$, and $13$ states, respectively.
The total $B_{\mathrm{GT}}$ from the $I=12$ mother state is $9.02$, which is $25\%$ smaller than the result of the 4-qp configuration in the single-$j$ limit with $j=7/2$.
It should be noted that the dominance of $I'=I-1$ transitions in the shell-model study is consistent with that of the  4-qp configuration in the deformed potential.

It is also noteworthy that the energy ordering of different $I'$ components provides information on the strength and sign of the spin-spin coupling.
Since the highest spin states are the energetically lowest, it can be speculated that attractive neutron-proton spin-spin coupling lowers the states with the maximum aligned configuration with $|\bm{S}|=|\bm{s}_1+\bm{s}_2|=1$.
Note that the $I'^\pi=13^+$, $12^+$, and $11^+ $ states have dominant configurations $(\nu f_{7/2} \pi f_{7/2})^{J^+} (f_{7/2}^2)_{\pi}^{6^+}$ with $J=7$, $6$, and $5$, respectively.
The speculation is consistent with the effective interaction adopted in the shell-model calculations.
The two-body neutron-proton matrix element $\langle f_{7/2}^2; J |V | f_{7/2}^2; J \rangle$ is the most attractive for the aligned configuration $|\bm{J}|=|\bm{L}+\bm{S}|=7$ ($L=6$ and $S=1$) with a value of $-2.666$, and is also attractive but to a lesser extent for $J=5$ with a value of $-0.784$.
However, it is repulsive for $J=6$ with a value of $0.278$.

The lower panel of Fig.~\ref{fig:fe52-gt} shows the GT transition strength distributions from the $0^+$ ground state in $^{52}$Fe calculated in the same manner.
In this case, the final states are all $1^+$ states, and the summed transition strength is $8.09$.

The comparison of these two GT transition strength distributions is very interesting.
On the one hand, around $70\%$ of the sum-rule strength in the case of transitions from the HSI state is concentrated in a narrow bump with a Lorentzian shape at the high excitation-energy region around $E_x=12$~MeV.
This indicates that the strong repulsive residual interactions in the $ph$ excitations create a collective state.
On the other hand, the strength distribution for the transitions from the ground state has a bump in the lower energy region around $E_x=5$~MeV.
Its width is about twice that of the transitions from the HSI state.
This implies that the role of the repulsive residual interactions is less significant.
Therefore, it can be said that the collectivity of the GT transitions from the HSI state is much stronger than
that of the GT transitions from the ground state from the point of view of the total transition strength and the corresponding width.

\section{$I^{\pi}=21^+$ isomer in $^{94}$Ag}\label{sec:Ag94}

Nucleus $^{94}$Ag also exhibits an isomeric state at $E=6.67$~MeV with half-life $T_{1/2}=0.40(4)$~s, while its $\gamma$-decay scheme is poorly known \cite{ENSDF}.
This state is considered to have spin-parity $I^{\pi}=21^+$.
The quasi-particle configuration is assigned as fully aligned the 6-qp states with the $[404\frac{9}{2}]_{\pi(\nu)}$, $[413\frac{7}{2}]_{\pi(\nu)}$, and $[422\frac{5}{2}]_{\pi(\nu)}$ orbits in the Nilsson diagram.
The GT matrix elements~(\ref{mem}), (\ref{me0}), and (\ref{mep}) in the intrinsic frame are
\begin{align}\label{int-me2}
  {2(6j-3)}/{j} \qquad &\mbox{for} \qquad \Delta K=-1\,,  \nonumber \\
  6 \qquad &\mbox{for} \qquad  \Delta K=0\,, \nonumber \\
  {6}/{j} \qquad &\mbox{for} \qquad \Delta K=+1\,,
\end{align}
with $j=9/2$, respectively.
The values in Eq.~(\ref{int-me2}) are obtained by assuming that the wave functions $[Nn_3\Lambda\Omega]=[404 \frac{9}{2}]_{\pi(\nu)}$, $[413 \frac{7}{2}]_{\pi(\nu)}$, and $[422 \frac{5}{2}]_{\pi(\nu)}$
have dominant contributions of $|j\Omega\rangle = |\frac{9}{2}\frac{9}{2}\rangle$, $|\frac{9}{2}\frac{7}{2}\rangle$, and $|\frac{9}{2}\frac{5}{2}\rangle$ orbits, respectively.

\begin{table}
  \caption{$B_{\mathrm{GT}}$ strengths for the transition $I\rightarrow I'$ with the 6-qp configuration of protons and neurons  $[Nn_3\Lambda\Omega]=[N0N(\Lambda+1/2)]_{\nu(\pi)}$, $[N1(N-1)(\Lambda+1/2)]_{\nu(\pi)}$, and $[N2(N-2)(\Lambda+1/2)]_{\nu(\pi)}$.  The sum values in the last line are evaluated with $I=21$ and $j=9/2$.}\label{tab-GT2}
\begin{ruledtabular}
 \begin{tabular}{cccc}
      &  $I'=I-1$    &  $I'=I$  & $I'=I+1$  \\\hline
      $\Delta K=-1$  &  $\frac{2(2I-1)}{2I+1}\frac{6j-3}{j}$ &     $\frac{2}{I+1}\frac{6j-3}{j}$ & $\frac{2}{(2I+1)(I+1)}\frac{6j-3}{j}$  \\
      $\Delta K=0   $   &---  &  $\frac{6I}{I+1}$ & $\frac{6}{I+1}$   \\
      $\Delta K=+1$¡¡   & ---   &   ---& $\frac{6}{j}$ \\\hline
      sum  & 10.17 & 6.21  & 1.61  \\
 \end{tabular}
\end{ruledtabular}
\end{table}

Table~\ref{tab-GT2} lists the non-energy-weighted summed GT transition strengths for different $I'$ final states.
The GT transition of $I\rightarrow I'=I-1$ is about $56.5\%$ of the total strength $B_{\mathrm{GT}}=18$.
The transition of $I\rightarrow I'=I$ has around $34.5\%$ of the total strength, whereas that of $I\rightarrow I'=I+1$ is rather weak with about $8.9\%$ of the total strength.
The ratios to the total sum-rule strength for different $I'$ states is quite similar for both $^{52}$Fe and $^{94}$Ag.

\begin{figure}
\begin{center}
    \includegraphics[width=8cm]{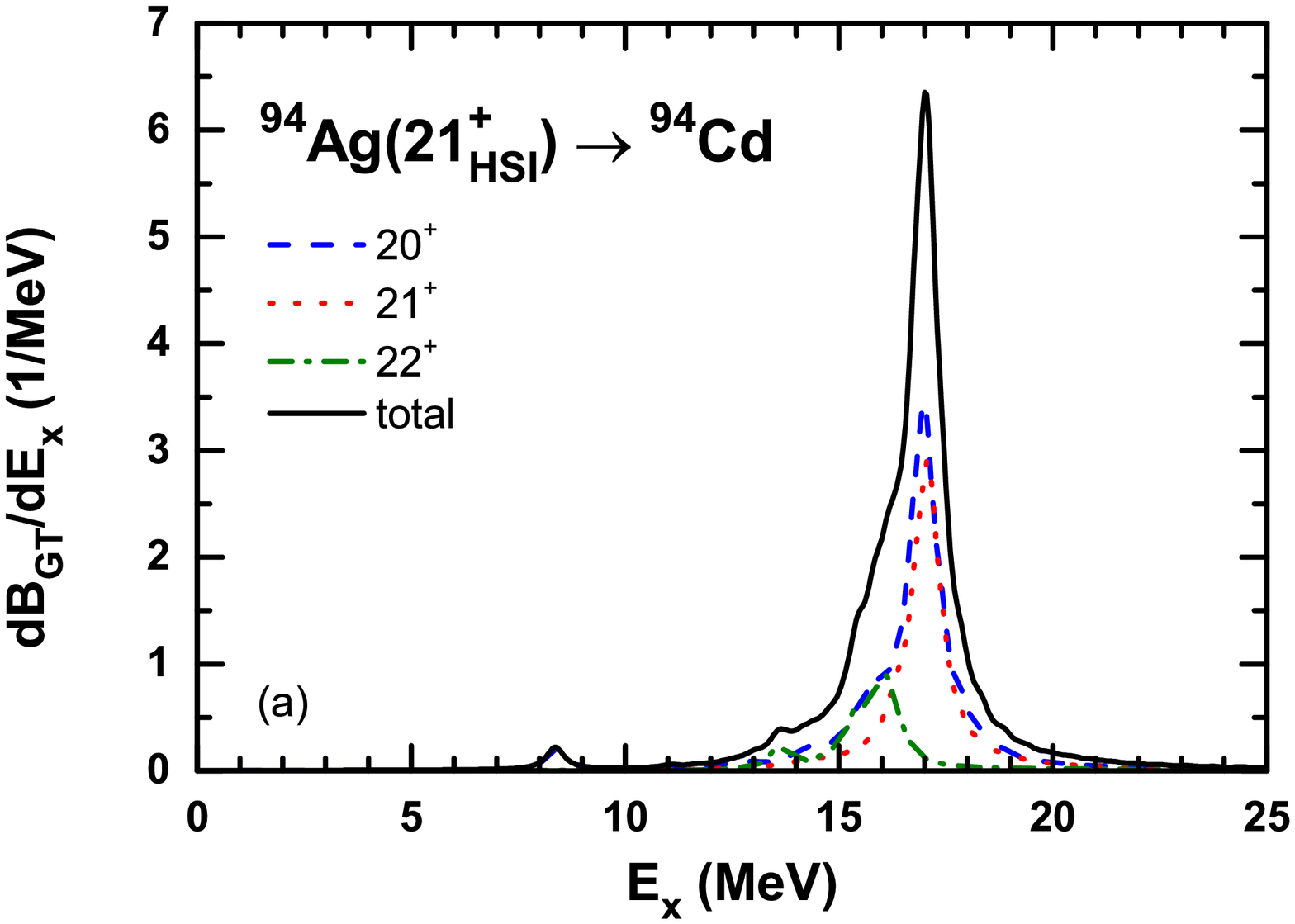}\\
    \includegraphics[width=8cm]{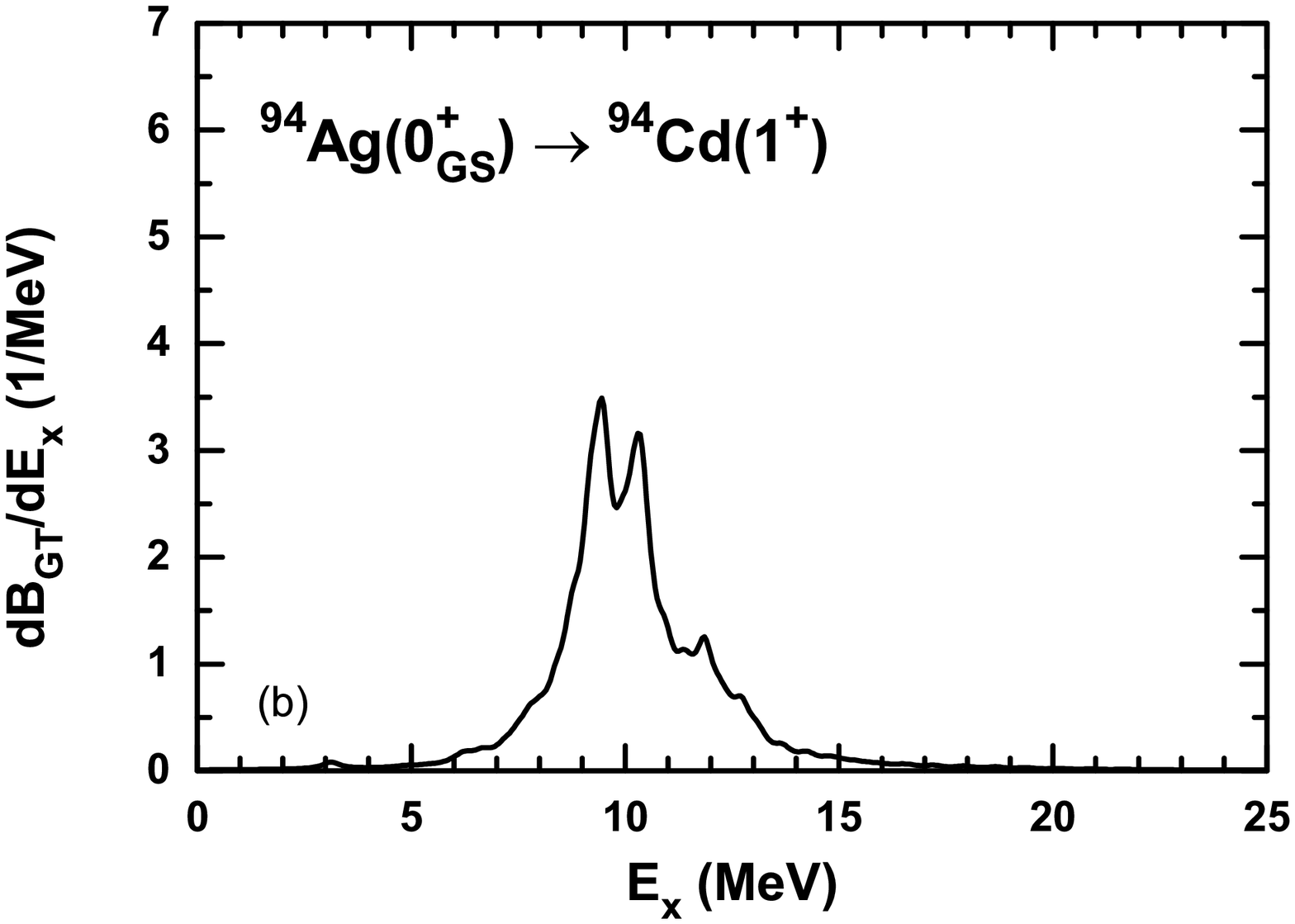}
    \caption{(Color online) (Upper) GT transition strength distributions from the high-spin isomeric state $I^{\pi}=21^+$ in $^{94}$Ag calculated by the shell model in the $(2p_{1/2},\, 1g_{9/2},\, 1g_{7/2},\, 2d_{5/2})$-shell model space with a modified P1GD5G3 \cite{Honma2014} effective interaction.
    Transition strengths are averaged by a Lorentzian weighting function with a width of $0.5$~MeV.
    Excitation energy is defined from the ground state of the daughter nucleus $^{94}$Cd.
    No quenching factor is adopted here.
    (Lower) Same as the top panel but for the GT transition strength distributions from the ground-state $I^{\pi}=0^+$ in $^{94}$Ag.
\label{fig:ag94-gt}}
\end{center}
\end{figure}

We perform shell-model calculations for the GT transitions from the $I^{\pi}=21^+$ HSI state in $^{94}$Ag.
The model space is taken as $(2p_{1/2},\, 1g_{9/2},\, 1g_{7/2},\, 2d_{5/2})$ with the $^{76}_{38}\mathrm{Sr}_{38}$ core.
We adopt a modified P1GD5G3 shell-model interaction for this study.
The P1GD5G3 \cite{Honma2014} interaction was originally optimized by fitting the experimental data in the mass range of $47<N<58$ starting from the G-matrix interaction derived from the chiral N$^3$LO interaction \cite{Entem2003Phys.Rev.C68_041001}.
For the present calculations, this interaction is further tuned by taking more experimental information into account.
The $I^{\pi}=21^+$ isomeric state is predicted at $E=6.79$~MeV, which is slightly higher than its experimental value $6.67$~MeV.
The shell-model configurations of the $I^{\pi}=21^+$ state are dominated by $74\%$ of $(g_{9/2}^3)_{\nu}(g_{9/2}^3)_{\pi}$ configuration with $7.4\%$ mixing of $(g_{9/2}^3)_{\nu}(g_{9/2}^2d_{5/2})_{\pi}$ or $(g_{9/2}^2d_{5/2})_{\nu}(g_{9/2}^3)_{\pi}$ as well as $5.0\%$ mixing of $(g_{9/2}^3)_{\nu}(g_{9/2}^2g_{7/2})_{\pi}$ or $(g_{9/2}^2g_{7/2})_{\nu}(g_{9/2}^3)_{\pi}$.
For the decay scheme, the present shell-model calculations show this $21^+$ state may decay to two $17^+$ states at $5.88$ and $6.46$~MeV by the $E4$ transitions which are highly forbidden.
Meanwhile, one keeps in mind that the decay branch from this HSI is dominated by the electron capture and $\beta^+$ decay \cite{ENSDF}.

The upper panel of Fig.~\ref{fig:ag94-gt} shows the calculated GT transition strength distributions.
The main strengths are found at around $E_{x}=17$~MeV for both the $I'^\pi=20^+$ and $21^+$ states.
A smaller peak is observed around $E_{x}=16$~MeV for the $I'^\pi=22^+$ state (i.e., $1$~MeV lower than the other two components).
The total $B_{\mathrm{GT}}$ is around $10.9$, which corresponds to around $61\%$ of the sum-rule strength of the 6-qp configuration in Table~\ref{tab-GT2}.
Such a small value may be due to the substantial configuration mixing (more than $20\%$) besides the $(g_{9/2}^3)_{\nu}(g_{9/2}^3)_{\pi}$ configuration in the $I^{\pi}=21^+$ state in $^{94}$Ag.
Compared to the total GT transition strength, the transitions to the different final states $I'^\pi=20^+,\,21^+$, and $22^+$ are $49.8\%$, $35.6\%$, and $14.7\%$, respectively.
These ratios are similar to those of sum-rule results shown in Table~\ref{tab-GT2}.

In this or a heavier mass region, the collectivity generally becomes strong due to the large number of nucleons involved in the excitations.
Similar to the case of $^{52}$Fe, the GT strength distribution in $^{94}$Ag from the HSI state has a narrower resonance structure at a high energy around $E_x=17$~MeV, compared with the GT strength distribution from the ground state, as shown in the lower panel of Fig.~\ref{fig:ag94-gt}.
Quantitatively, $90\%$ of the total strength is found within about a $5$~MeV energy window for the GT transitions from the isomeric state, while the energy window is about $7$~MeV for the transitions from the ground state.

The results of both $^{52}$Fe and $^{94}$Ag support the simplified view explained in the introduction.
The GT transitions from the HSI states exhibit a strong collectivity due to the strong $ph$ residual interactions in the asymmetric Fermi spheres of the spin-up and spin-down components.
On the other hand, the transitions from the $0^+$ ground states of $N=Z$ nuclei are due to more complex excitations.
Consequently, the strength distribution cannot be understood only through simple $ph$ excitations.
In other words, the Fermi spheres of the initial ground states are melted.
These specific features can be seen through the measurements of the GT transitions from these two states.

\section{Summary}\label{sec:Summary}

We derive a new sum rule for the GT transitions from the high-spin isomeric states in $N=Z$ nuclei.
The shell-model calculations with modern effective interactions are also performed to study the GT transitions from both the ground states and the HSI states in $^{52}$Fe and $^{94}$Ag.
It is pointed out that the GT strengths from the HSI states show a stronger collectivity than those from the ground states.
To understand the physical mechanism of this enhancement, the concept of spin Fermi spheres, which are asymmetric for the spin-up and spin-down components in the HSI states, is important, as it plays an analogous role to the isospin Fermi spheres in the $N>Z$ nuclei to create a large collectivity.

\begin{acknowledgments}
This work was supported, in part, by JSPS KAKENHI under Grants No.~16K05367, No.~15K05090, and No.~18K13549, and the RIKEN iTHES Project and iTHEMS Program.
\end{acknowledgments}


\end{document}